# Hidden Orbital Liquid State Within Ferromagnetically Ordered Metallic SrRuO$_3$

I.M. Bradarić,[1,2] M.S. Laad,[3] F.V. Kusmartsev,[1] K. Yoshii,[4] and S. Okayasu[5]

[1] *Department of Physics, Loughborough University, Loughborough, LE11 3TU, United Kingdom*
[2] *Laboratory for Theoretical and Condensed Matter Physics, The Institute of Nuclear Sciences "Vinča", P.O. Box 522, 11001 Belgrade, Serbia*
[3] *Max-Planck-Institut für Physik Komplexer Systeme, 01187 Dresden, Germany*
[4] *Quantum Beam Science Directorate, Japan Atomic Energy Agency, Hyogo 679-5148, Japan*
[5] *Advanced Science Research Center, Japan Atomic Energy Agency, Ibaraki 319-1195, Japan*

We have experimentally found related anomalies in electrical resistivity, dc and ac magnetic susceptibility, appearing deeply within ferromagnetically ordered state in SrRuO$_3$. Lack of Jahn-Teller distortion in this regime rules out conventional orbital order, forcing one to describe these in terms of an orbital liquid ground state coexisting with ferromagnetic spin order. We suggest that weak spin-orbit coupling in such an unusual state underpins the observed anomalies.

PACS numbers: 71.28.+d, 71.70.Ej, 75.30.Cr, 72.80.Ga

Complexities in the ground state properties of the transition-metal oxides, and especially in compounds that are close to the either side of metal-to-insulator transition, remain one of the central issues in condensed matter physics. These complexities arise due to competing interactions in charge, spin, orbital, and lattice degrees of freedom, where subtle changes in material composition, pressure, magnetic field etc., may tip the balance to a plethora of exotic ground states, like unconventional high temperature superconductivity, and colossal magneto-resistance. The Ruddlesden-Popper (RP) [1] series of ruthenates Sr$_{n+1}$Ru$_n$O$_{3n+1}$, where $n$ denotes the number of Ru-O layers separated by Sr-O layers, offer a unique opportunity to study the balance between various degrees of freedom with varying $n$ (i.e., effective dimensionality). In these systems interaction between Ru atoms is realized via oxygen atoms (corner sharing RuO$_6$ octahedra) through strong hybridization of Ru t$_{2g}$ and O 2p orbitals. Therefore, the octahedral cubic crystal field splits Ru$^{4+}$ d-band into a lower threefold degenerate t$_{2g}$ level, and a higher energy, empty twofold degenerate e$_g$ level, resulting in a low spin configuration Ru$^{4+}$(t$_{2g}^4$,e$_g^0$: S=1) [2]. In the resulting orbital degenerate cases [3], strong orbital fluctuations couple to spins via spin-orbit coupling, making ruthenates attractive candidates in the search for exotic ordered states of matter.

Apparently, unconventional superconducting ground state (p-wave, spin triplet symmetry) in quasi-two-dimensional Sr$_2$RuO$_4$ shows *orbital dependency* [4]. The second member Sr$_3$Ru$_2$O$_7$, is a paramagnetic Fermi liquid with strongly enhanced magnetic susceptibility and ferromagnetic (FM) instability at low temperatures, which undergoes FM phase transition upon applying a slight uniaxial pressure [5]. Furthermore, this system exhibits magnetic field-tuned quantum criticality, leading to quantum critical end point [6]. Importantly, the high magnetic field magnetization, magnetic torque [7], and magneto-resistivity experiments [8], suggest *orbital dependent metamagnetism* in this compound. The trilayered Sr$_4$Ru$_3$O$_{10}$ shows complex magnetic behavior comprising the interlayer FM below approximately 105 K, and intralayer metamagnetism [9]. Recent studies show strong coupling between the spin and lattice degrees of freedom [10], magnetic field induced quantum criticality [11], and complex *orbital dependent magnetism* including the electronic phase separation [12]. Clearly, these compounds display an appealing diversity in physical properties, involving crucial role of the orbital degrees of freedom, so that the compelling question naturally arises: What is the signature of the *orbital physics* in the infinite layered, nearly three dimensional, SrRuO$_3$?

The apparently conventional ferromagnet (T$_c$ = 163 K) SrRuO$_3$ shows anomalous electronic transport above 10 K [13], featuring $\rho \sim \sqrt{T}$ dependence above T$_c$ and violation of the Mott-Ioffe-Regel limit [14]. Furthermore, the real part of optical conductivity deviates sharply from the Landau's Fermi liquid behavior, with an anomalous power-law dependence on frequency, $\sigma \sim 1/\sqrt{\omega}$ [15]. Notably, magnetotransport results [16,17], and anomalous Hall effect [18], point to an unexpected and puzzling dynamics far below T$_c$, observed in single crystals, single crystal thin films, and ceramic samples. These observations motivated us to closely examine magnetic response of SrRuO$_3$ within intermediate temperature range. In this Letter we present results of ac and dc magnetic susceptibility, and electrical resistivity. We find hitherto unreported, but closely linked, anomalies in $\rho(T)$, M(T), and a broad peak in the real part of the ac magnetic susceptibility, *all* centered at ~61 K ($f$=10$^{-2}$ Hz). The latter shifts to lower temperatures with increasing frequency of the ac driving field. We show how these anomalies are consistently reconciled *only* within an underlying picture of strong orbital fluctuations.

Polycrystalline samples were synthesized starting with high purity (99.999+) SrCO$_3$ and RuO$_2$ in ratio 1:1.02.The mixed materials were prefired at $820^0$ C for total of 100 hours, and then repeatedly sintered at $1100^0$ C – $1300^0$ C for another 100 hours in air. X-ray diffraction and dc magnetization measurements verified phase pure $SrRuO_3$. Magnetic measurements were performed with Quantum Design MPMS XL SQUID magnetometer and PPMS. All ac magnetic susceptibility measurements were performed using driving field $H_{ac}$ = 4 Oe. Electrical resistivity measurements were performed using standard four probe method.

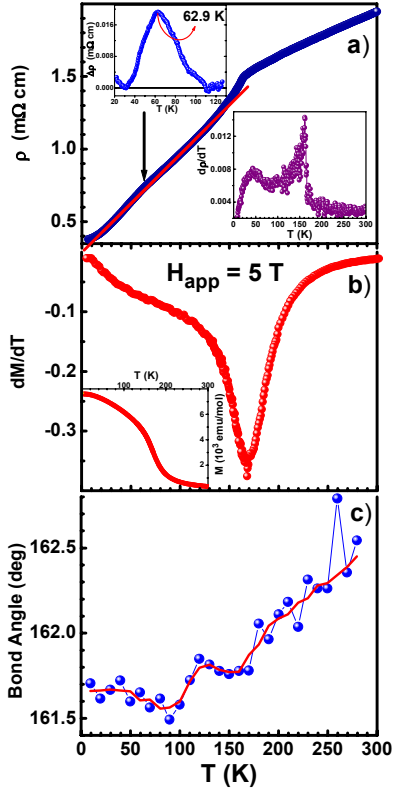

FIG. 1 (a) The electrical resistivity vs. temperature. The red line is a linear function drawn between 30 K and 110 K; upper inset shows difference between $\rho(T)$ and the linear function, and lower inset shows temperature derivative of $\rho(T)$. (b) The temperature derivative of magnetization and M(T) in the inset. (c) The temperature dependence of the bond angle Ru-$O_1$-Ru, taken from Ref. 19. Red line is smoothed curve using fast Fourier transform.

Figure 1a shows resistivity vs. temperature without applied magnetic field. Aside from the well known decrease of resistivity at $T_c$ (originating from a decrease of the charge scattering rate by spin fluctuations), Fermi liquid behavior below ~11 K, and $\sqrt{T}$ dependence above $T_c$, a broad hump is visible far below $T_c$. When a linear function (drawn between 30 K and 110 K), is subtracted from $\rho(T)$ (upper inset in Fig. 1a), a sharp peak at ~63 K appears. Simultaneously, $d\rho(T)/dT$ curve (lower inset in Fig. 1a) shows sudden increase at ~63 K, reaching maximum (the fastest decrease in $\rho(T)$) at ~43 K. This behavior clearly demonstrates that an additional mechanism, enhancing the charge scattering rate, is present in the system (*hidden order*). Corresponding anomalous behavior is apparent in dM(T)/dT curve shown in Fig. 1b., where change of curvature (below $T_c$) is observed around 130 K and 60 K. Clearly, classical spin-wave excitations cannot account for this anomaly, since the fitting of the Bloch's law (not shown) to M(T) failed for all temperature ranges. Furthermore, Ru-$O_1$-Ru bond angle dependence on temperature (Fig. 1c [19]), shows continuous decrease from room temperature to $T_c$, followed by a small downward step between ~130 K and ~90 K. This means that conduction band is slightly reduced, which is reflected in slower decrease of resistivity in this temperature range, and a small change in magnetocrystalline anisotropy due to spin-orbit interaction. Since spin-orbit coupling (SOC) constant is temperature independent, the only possible source of the observed anomalous behavior lies in the orbital sector. This should not be surprising (for nearly cubic perovskite structure of $SrRuO_3$ with high crystal field symmetry), since the threefold degeneracy of the $t_{2g}$ orbitals gives rise to strong orbital frustration, leading to large quantum disorder in the orbital sector, i.e., *quantum orbital liquid* ground state. This issue has been recently theoretically explored to account for the non Fermi liquid behavior observed in $SrRuO_3$ and $CaRuO_3$ [20]. Importantly, neutron diffraction studies [21] and a magnetic Compton-profile studies [2] proved the absence of the Jahn-Teller distortion (orbital-lattice coupling), which would lift the orbital degeneracy, and consequently, trigger orbital ordering. On the other hand, SOC constant $\lambda$=0.161 eV [22] is much less than conduction bandwidth W=2.6 eV [23], and thus, would induce only *small* orbital splitting that could be easily wiped out by carrier itinerancy. Consequently, orbital fluctuations survive at all temperatures: this appears to be crucial for the metallic behavior of the system, as discussed in Ref. 24. However, finite SOC introduces a small orbital mixing and therefore effectively reduces the orbital degeneracy, leading to the dynamical spin-orbital entanglement. This means that spin fluctuations will introduce additional quantum disorder in the orbital sector, possibly preventing weak orbital order (which might have been expected due to deviations of the bond angles from $^o$180 of ideal perovskite symmetry). Of course, orbital fluctuations will, vice versa, introduce disorder in the spin sector, which is exactly the nature of the anomalous behavior described above.

In this spirit, we examined dynamical magnetic response of the system, varying the frequency of the ac driving magnetic field (Fig. 2a), and the magnitude of the applied dc magnetic field (Fig. 3a,b). Noticeable features are (upper part of Fig. 2a, and upper inset), a sharp kink in the real part ($\chi'$) of the ac magnetic susceptibility at ~130 K, and a broad peak centered at 59.3 K for $f = 0.1$ Hz (reaching 61 K for the dc limit $f \sim 0.01$ Hz), perfectly matching the anomalies found in $\rho(T), M(T)$, and bond angle vs. temperature, above.

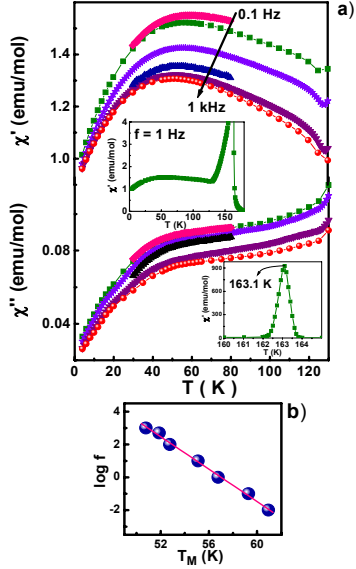

FIG. 2. (a) The frequency (from the top 0.1, 1, 10, 100, 500, 1000 Hz) and temperature dependence of: $\chi'$, upper part and $\chi''$, lower part. The upper inset shows $\chi'$ ($f=1$ Hz) at larger temperature range, and lower inset depicts the sharp peak of $\chi'$ at $T_c$. Noticeable is a large ratio $\chi'(T_C,1$ Hz$) / \chi'(T_M,1$ Hz$) = 411$. (b) The logarithm of frequency vs. $T_M$ (temperature position of the broad maximum in $\chi'$).

The exciting feature is a decrease of $T_M$ (the temperature position of the maximum in $\chi'$) with increasing $f$, depicted in Fig 2b., having functional form $ln(f/f_0) = -kT_M$. This kind of frequency dependence signals presence of the multiple relaxation rates in the system, as found in $NaNiO_2$ above antiferromagnetic ordering temperature [25]. In the orbital liquid picture [20], such multiple relaxation rates can readily arise from degenerate, dynamical, orbital configurations coupled to spin dynamics. This is a clear evidence of the dominant role of quantum orbital fluctuations (with pseudospin T = 1/2) rather than by the large spins S = 1, which would behave semiclassically. In this sense, $T_M$ marks crossover temperature below which quantum orbital fluctuations become dominant over thermal fluctuations. We recall here that orbital dynamics is reflected in magnetic fluctuations only indirectly, via SOC. Consistently, $\chi'(T)$ curves converge approaching T = 0 K, implying that, at low T, only low energy spin fluctuations are thermally excited, and quantum orbital fluctuations dominate, so that the signal is attenuated and almost frequency independent. This also explains why the full saturation moment 2 $\mu_B$/Ru is not reached even in 44 T magnetic fields at 4.2 K [26]; quantum orbital fluctuations depress the magnetization at low T.

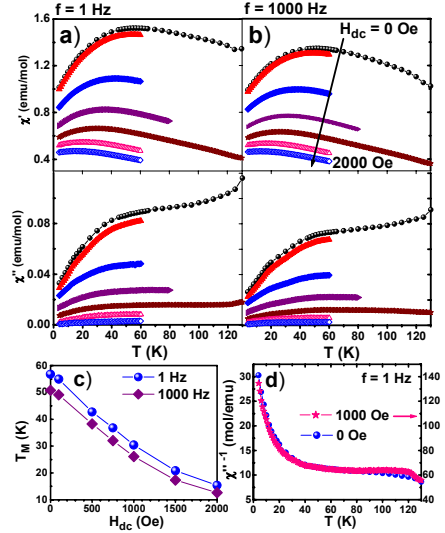

FIG. 3. The dependence on the applied dc magnetic field (from the top 0, 100, 500, 750, 1000, 1500, and 2000 Oe) of $\chi'$, upper part and $\chi''$, lower part, for $f=1$ Hz (a), and $f=1000$ Hz (b). (c) The dependence of $T_M$ on applied dc magnetic field, for $f = 1$, and 1000 Hz. (d) The temperature dependence of the inverse of imaginary susceptibility for $H_{dc} = 0$, and 1000 Oe, at $f = 1$ Hz.

The influence of the applied dc magnetic field (shown in Fig 3a, and 3b for $f=1$ Hz and $f=1000$ Hz, respectively) is also very interesting. Clearly, $T_M$ and the difference between $T_M$ ($f = 1$ Hz) and $T_M$ ($f = 1$ kHz), decrease with increasing $H_{dc}$ (Fig 3c). As discussed above, dc magnetic field depresses spin fluctuations (i.e., has the same effect as decreasing temperature), which results in decreasing $T_M$. Accordingly, $T_M(f,H)$ curves converge and saturate to a constant value at lower temperatures, where the quantum-regime is dominant. Unfortunately, as the signal progressively becomes very low, it was not possible to accurately determine $T_M$ at larger H.

We notice that $\chi''(T)$ (lower part of the Fig. 2a) does not follow the behavior of $\chi'$, and shows the broad plateau above approximately 60 K. Furthermore, $H_{dc}$ does not affect the shape of $\chi''^{-1}$ (i.e., only the magnitude increases with increasing field, Fig 3d). Recall that $\chi''$ reflects energy dissipation of the ac magnetic field, which in this case means, induced eddy currents in the system, and therefore, $\chi''^{-1}$ is proportional to magnetoresistivity. Indeed, it shows qualitatively close similarity to negative magnetoresistance [17], and resistivity parallel to domain walls [16], which points to intrinsic microscopic mechanism of temperature dependence. Apparently, decrease in negative magnetoresistance [17], and resistivity below 63 K, reflects dominant scattering of charge carriers by orbital fluctuations, caused perhaps by spin-orbit coupling becoming relevant. Eventually, in the low-T (< 10 K) correlated Fermi liquid regime, electrical resistivity is entirely governed by quantum orbital fluctuations, since long wavelength magnons have negligible influence. Here resistivity dependence on magnetic field enters only through SOC. Instructively, resistivity anisotropy deduced from planar Hall effect (Fig. 4 in Ref. 27) is nonzero at all temperatures, and shows sudden increase on approaching $T_c$ from the ordered side, followed by a slow decrease above $T_c$. In a renormalized Fermi liquid (FL) picture, this could arise from anisotropic Fermi velocities in the paramagnetic state; however, given the non-FL behaviour found in this T-regime, this is unlikely, since even the Luttinger Fermi surface is not well-defined above T=10 K. Moreover, there is another large decrease in anisotropy below ~65 K, coinciding with increase in $\chi''^{-1}$. Evidently, this behavior is not related to spontaneous magnetization, and obviously, for the same reason as above (T > 10 K) cannot be explained considering anisotropic Fermi velocities arising from the orthorhombic distortion of the lattice. Alternatively, the theoretical model developed in Ref. 20, predicts separate Fermi velocities for charge and orbital-density waves (i.e., charge-orbital pseudospin separation), explaining non-FL properties in $CaRuO_3$ and $SrRuO_3$. In such an orbital liquid, SOC will introduce anisotropy in the velocities of the orbital modes, now coupled to the magnetization. These will show up in the Hall data as a resistivity anisotropy, which is now a measure of the SOC. Seemingly, the system could be close to a dynamical, orbital ordered state as temperature is lowered below ~130 K. However, strong quantum orbital fluctuations prevail at all T, rendering the system an incoherent metal [20], and suppressing the emergence of a FL state to T<10 K. Given absence of JT distortions, conventional orbital order is ruled out. Whether SOC in a triply degenerate orbital setting can induce exotic orbital ordered states (they will involve breaking of time-reversal symmetry) of the orbital current type is an interesting question deserving of further study.

In conclusion, the interplay between spin and orbital degrees of freedom interacting via small SOC in $SrRuO_3$ is revealed in anomalous temperature dependence in static and dynamic magnetic susceptibility, and resistivity. This exposes the important role of $t_{2g}$ orbital degeneracy and its interplay with spin-orbit coupling in bad metallic 4d-shell based transition metal oxides with almost ideal perovskite based structure.

This research has been supported by Royal Society Grant No. 2004/R3-EF, Serbian Ministry of Science, Project No. 141014/2006, and ESF network AQDJJ. MSL thanks the MPIPKS, Dresden for financial support.